# Subspace System Identification via Weighted Nuclear Norm Optimization

Anders Hansson *
Division of Automatic Control
Linkoping University
SE–581 83 Linkoping, Sweden
anders.g.hansson@liu.se

Zhang Liu
Northrop Grumman Corporation
16710 Via Del Campo Court
San Diego, CA 92127, USA
zhang.liu@gmail.com

Lieven Vandenberghe
Electrical Engineering Department
UCLA, 66–147L Engineering IV
Los Angeles, CA 90095, USA
lieven.vandenberghe@ucla.edu

*Abstract*— We present a subspace system identification method based on weighted nuclear norm approximation. The weight matrices used in the nuclear norm minimization are the same weights as used in standard subspace identification methods. We show that the inclusion of the weights improves the performance in terms of fit on validation data. As a second benefit, the weights reduce the size of the optimization problems that need to be solved. Experimental results from randomly generated examples as well as from the Daisy benchmark collection are reported. The key to an efficient implementation is the use of the alternating direction method of multipliers to solve the optimization problem.

## I. INTRODUCTION

Subspace algorithms include some of the most popular methods for system identification [Lju99], [VV07]. One of the main reasons for their success is their reliance on efficient matrix algorithms (QR and singular value decomposition) for making low-rank approximations of matrices constructed from the observed inputs and outputs. By incorporating weights in the approximation problems one can also accommodate colored noise characteristics in the state-space model estimation. A well-known drawback of subspace methods is that they offer no guarantee of efficiency in the estimates. In particular, it is not known whether the Cramer-Rao lower bound is achieved. From a practical point of view the selection of the model order by thresholding the singular values can also be difficult.

Recently there has been some interest in nuclear norm minimization as an alternative technique for finding low-rank matrix approximations. Minimizing the nuclear norm (sum of singular values) is a popular convex heuristic for low rank matrix approximation, first proposed in [FHB01], [Faz02]. It offers the important advantage that it preserves linear matrix structure, unlike the SVD commonly used in subspace methods. Nuclear norm optimization is also easily combined with convex regularization terms and convex constraints on the parameters. Its use is further motivated by the spectacular success of closely related techniques in signal processing and machine learning, in particular, sparse optimization (1-norm methods) [CRT06], [Don06], [Tro06] and low-rank matrix completion [CR09], [RFP10], [CT10]. Experiments with nuclear norm formulations of simple subspace methods in [LV09a], [LV09b], [MF10], [GvWvdVV11] indicate that the nuclear norm heuristic can be quite effective in identification as well. However it should be noted that the results of the nuclear norm heuristic can be very suboptimal (in particular, when used for approximations with a fixed rank; see, for example, [Mar11]), and at present no theoretical analysis of its effectiveness for identification is available.

The purpose of this paper is to improve the algorithm in [LV09a] by generalizing it to state-space models in Kalman normal form and combining it with instrument variable techniques [Ver94], [OD94], [Vib95], [Lar90], [VWO97]. Experiments on randomly generated problems will show that the instrument variable based nuclear norm approach improves the fit in subspace identification in more than 90% of the cases considered, whereas the same figure for the case without instrument variables is only 84 %. Moreover, using instrument variables reduces the dimension of the optimization problem making it possible to obtain the solution faster. Hence instrument variables increase efficiency with respect to both speed and fit.

The paper is organized as follows. In Section 2 we review the basics of subspace identification, and in Section 3 we recapitulate the nuclear norm heuristics for minimizing the rank of a matrix subject to linear constraints. Then in Section 4 a simple alternating direction methods of multiplier (ADMM) algorithm for solving nuclear norm optimization problem with quadratic regularization is presented. In Section 5 the proposed approach is evaluated on examples. Finally, in Section 6 we make some concluding remarks.

## II. SUBSPACE IDENTIFICATION

We consider identification of the following linear discrete-time state-space model on Kalman normal form:

$$x(k+1) = Ax(k) + Bu(k) + Ke(k) \qquad (1)$$
$$y(k) = Cx(k) + Du(k) + e(k) \qquad (2)$$

where $x(k) \in \mathbf{R}^{n_x}$, $u(k) \in \mathbf{R}^{n_m}$, $e(k) \in \mathbf{R}^{n_p}$ and $y(k) \in \mathbf{R}^{n_p}$. It is assumed that $e(k)$ is ergodic, zero mean, white noise, see e.g. [Lju99]. The system matrices $A, B, C, D, K$ are real-valued. The objective is to estimate the system matrices from a sequence of observed inputs $u(k)$ and outputs $y(k)$.

Subspace identification is based on the following block

---

Hankel matrix of outputs

$$Y_{i,r,N} = \begin{bmatrix} y(i) & y(i+1) & \ldots & y(i+N-1) \\ y(i+1) & y(i+2) & \ldots & y(i+N) \\ \vdots & \vdots & \ddots & \vdots \\ y(i+r-1) & y(i+r) & \ldots & y(i+N+r-2) \end{bmatrix}, \quad (3)$$

and a similarly defined input block Hankel matrix $U_{i,r,N}$. When only output noise is considered (i.e., $K = 0$), the subspace identification scheme first forms the matrix

$$G = \frac{1}{N} Y_{0,r,N} \Pi^{\perp}_{U_{0,r,N}} \quad (4)$$

where

$$\Pi^{\perp}_{U_{0,r,N}} = I - U^T_{0,r,N}(U_{0,r,N}U^T_{0,r,N})^{-1}U_{0,r,N}$$

is a projection matrix on the nullspace of $U_{0,r,N}$. An SVD of $G$ is computed to estimate the state-order $n_x$ of the dynamical system and make a low rank approximation of $G$. Based on the truncated SVD straightforward computations are performed to determine $A, B, C, D$ [Lju99].

For the case of colored noise, instrument variables may be used to avoid biased estimates; see e.g. [Lju99]. This results in a simple modification of the matrix $G$ to

$$G = \frac{1}{N} Y_{s,r,N} \Pi^{\perp}_{U_{s,r,N}} \Phi^T \quad (5)$$

where

$$\Phi = \begin{bmatrix} U_{0,s,N} \\ Y_{0,s,N} \end{bmatrix} \quad (6)$$

is the so-called instrument variable. Several different variations of this method exist, which can all be described as weighting modifications of $G$ as

$$\hat{G} = W_1 G W_2. \quad (7)$$

The weight matrices $W_1$ and $W_2$ for the different methods are given in Table I. (Note that in [Lju99] the time indexes are shifted and the rows in $\Phi$ are permuted as compared to the notation we are using, but this does not make any difference, since the singular values are independent of these permutations.) We remark that the dimension of the matrix $\hat{G}$ depends on the weighting as can be seen in the last row of the table. If no weighting is used, the dimension of $\hat{G}$ is $r \times 2s$. If no instrument variables are used the dimension is $r \times N$. Typically $s$ is much smaller than $N$. Hence the IVM method, CVA method, and instrument variable method with no weighting involve matrices with the lowest dimension, whereas the method with no instrument variables, MOESP, and N4SID have the highest dimension. This affects the computational time when minimizing the nuclear norm of $\hat{G}$.

## III. NUCLEAR NORM OPTIMIZATION

In this section we review the nuclear norm optimization problem and discuss how it can be used as a heuristic for low-rank approximations of structured matrices.

The nuclear norm $\|X\|_*$ of a matrix $X \in \mathbf{R}^{p \times q}$ is defined as the sum of the singular values of $X$. We will be interested in minimizing the nuclear norm of a matrix that depends affinely on some vector $\mathbf{x} \in \mathbf{R}^n$, i.e.

$$\text{minimize } \|\mathcal{A}(\mathbf{x}) - \mathcal{B}\|_*. \quad (8)$$

This is popular as a convex heuristic for

$$\text{minimize } \mathbf{rank}(\mathcal{A}(\mathbf{x}) - \mathcal{B}). \quad (9)$$

The convex nuclear norm heuristic for matrix rank minimization was first proposed by Fazel, Hindi, and Boyd in [FHB01], and is motivated by the observation that the solution typically has low rank.

The idea was applied to subspace system identification in [LV09a]. The method of [LV09a] is based on solving a regularized approximation problem

$$\text{minimize } \|G(\mathbf{y})\|_* + \lambda \|\mathbf{y} - \mathbf{y}_{\text{meas}}\|_2^2 \quad (10)$$

where $\lambda$ is a positive constant. Here

$$\mathbf{y}^T_{\text{meas}} = \begin{bmatrix} y_{\text{meas}}(0^T) \ldots y_{\text{meas}}(N+r-2)^T \end{bmatrix}$$

is the measured output sequence and $\mathbf{y}$ is an optimization variable of the same dimension as $\mathbf{y}_{\text{meas}}$. We define $\mathbf{u}_{\text{meas}}$ similarly. The matrix $G(\mathbf{y})$ is defined in (4) with the output Hankel matrix constructed from the optimization variable $\mathbf{y}$.

In this paper we extend this method to include instrument variables and matrix weights. Instead of minimizing the nuclear norm of $G$ in (4) we minimize the nuclear norm of $\hat{G}$ in (7). The variables are the outputs that define the Hankel matrix $Y_{s,r,N}$. However, we use the measured output when defining the instrument variable $\Phi$ and the weightings $W_1$ and $W_2$ in $\hat{G}$, in order to make $\hat{G}$ affine in $\mathbf{y}$. One can interpret the optimization problem as trying to find a new output $\mathbf{y}$ which will minimize the nuclear norm of $\hat{G}(\mathbf{y})$ at the same time as not deviating too much from $\mathbf{y}_{\text{meas}}$. To summarize we solve

$$\text{minimize } \|\hat{G}(\mathbf{y})\|_* + \lambda \|\mathbf{y} - \mathbf{y}_{\text{meas}}\|_2^2 \quad (11)$$

where $\hat{G}$ is defined via (5)–(7), where $Y_{s,r,N}$ is defined to be a function of the variable $\mathbf{y}$ and where all other matrices are defined via the measured output $\mathbf{y}_{\text{meas}}$ and the measured input $\mathbf{u}_{\text{meas}}$.

## IV. ADMM ALGORITHM

The regularized nuclear norm optimization problem (10) can be expressed as

$$\text{minimize } \|\mathcal{A}(\mathbf{x}) - \mathcal{B}\|_* + (1/2)(\mathbf{x} - \mathbf{x}_0)^T \mathcal{C}(\mathbf{x} - \mathbf{x}_0). \quad (12)$$

The variables are $\mathbf{x} \in \mathbf{R}^n$. The problem dimensions are defined by the linear mapping $\mathcal{A} : \mathbf{R}^n \to \mathbf{R}^{p \times q}$,

$$\mathcal{A}(\mathbf{x}) = A_1 \mathbf{x}_1 + A_2 \mathbf{x}_2 + \cdots + A_n \mathbf{x}_n.$$

The matrix $\mathcal{C} \in \mathbf{R}^{n \times n}$ is positive semidefinite. Algorithms for solving (12) were investigated in [LV09a], where a customized interior-point method was proposed. Since then, a number of first order methods appeared in [MF10], [LV10], [FPST12], including the accelerated gradient projection, alternating direction methods of multipliers (ADMM), and proximal point algorithms applied to the primal or dual

TABLE I
DEFINITIONS OF THE WEIGHT MATRICES IN DIFFERENT SUBSPACE IDENTIFICATION METHODS AND THE DIMENSIONS $p \times q$ OF $\hat{G}$.

| Method | MOESP | N4SID | IVM | CVA |
|---|---|---|---|---|
| $W_1$ | $I$ | $I$ | $\left(\frac{1}{N}Y_{s,r,N}\Pi^{\perp}_{U_{s,r,N}}Y_{s,r,N}^T\right)^{-1/2}$ | $\left(\frac{1}{N}Y_{s,r,N}\Pi^{\perp}_{U_{s,r,N}}Y_{s,r,N}^T\right)^{-1/2}$ |
| $W_2$ | $\left(\frac{1}{N}\Phi\Pi^{\perp}_{U_{s,r,N}}\Phi^T\right)^{-1}\Phi\Pi^{\perp}_{U_{s,r,N}}$ | $\left(\frac{1}{N}\Phi\Pi^{\perp}_{U_{s,r,N}}\Phi^T\right)^{-1}\Phi$ | $\left(\frac{1}{N}\Phi\Phi^T\right)^{-1/2}$ | $\left(\frac{1}{N}\Phi\Pi^{\perp}_{U_{s,r,N}}\Phi^T\right)^{-1/2}$ |
| $(p,q)$ | $(r,N)$ | $(r,N)$ | $(r,2s)$ | $(r,2s)$ |

TABLE II
TABLE SUMMARIZING THE ADMM ALGORITHM.

| Summary of the ADMM algorithm |
|---|
| 1. Initialize $\mathbf{x}$, $X$, $Z$, $t$. For example, $\mathbf{x} = 0$, $X = -\mathcal{B}$, $Z = 0$, $t = 1$. |
| 2. Compute factorization of $\mathcal{C} + tM$ |
| 3. Update $\mathbf{x}$ using (13) |
| 4. Update $X$ using (14) |
| 5. Update $Z := Z + t(\mathcal{A}(\mathbf{x}) - X - \mathcal{B})$ |
| 6. Check stopping criteria $\|r_\text{p}\|_F \leq \epsilon_\text{p}$ and $\|r_\text{d}\|_2 \leq \epsilon_\text{d}$. If not met, go to step 3. |

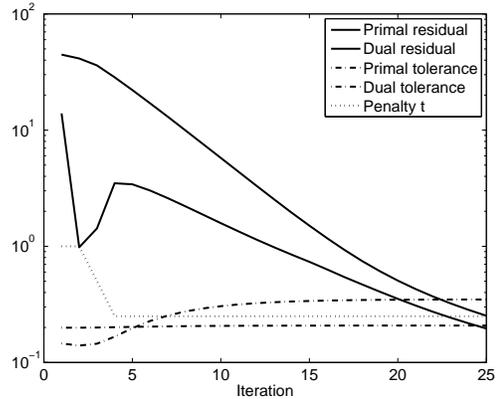

Fig. 1. Convergence of ADMM for a randomly generated problem with $p = q = 200$ and $n = 600$.

problem. In this section, we briefly describe the ADMM algorithm. For more details on ADMM, including experiments in system identification, we refer the reader to the survey papers [BPC+11], [FPST12]

We first write (12) as

$$\text{minimize} \quad \|X\|_* + (1/2)(\mathbf{x} - \mathbf{x}_0)^T \mathcal{C}(\mathbf{x} - \mathbf{x}_0)$$
$$\text{subject to} \quad \mathcal{A}(\mathbf{x}) - X = \mathcal{B}$$

and define the augmented Lagrangian

$$L_t(X, \mathbf{x}, Z) = \|X\|_* + (1/2)(\mathbf{x} - \mathbf{x}_0)^T \mathcal{C}(\mathbf{x} - \mathbf{x}_0) + \mathbf{Tr}(Z^T(\mathcal{A}(\mathbf{x}) - X - \mathcal{B})) + (t/2)\|\mathcal{A}(\mathbf{x}) - X - \mathcal{B}\|_F^2,$$

where $t > 0$ is a penalty parameter. If $t = 0$, this is the standard Lagrangian. Each iteration of ADMM in each iteration involves a minimization of $L_t$ over $\mathbf{x}$, a minimization of $L_t$ over $X$, and a simple update of the dual variable $Z$. The minimizer $\mathbf{x}$ can be obtained by simple differentiation:

$$\mathbf{x} = (\mathcal{C} + tM)^{-1}(\mathcal{A}_\text{adj}(tX + t\mathcal{B} - Z) + \mathcal{C}\mathbf{x}_0), \quad (13)$$

where $\mathcal{A}_\text{adj}$ is the adjoint of $\mathcal{A}$ and the matrix $M$ is defined by the identity $M\mathbf{x} = \mathcal{A}_\text{adj}(\mathcal{A}(\mathbf{x}))$. The minimizer $X$ can be obtained from singular value thresholding,

$$X = U \, \mathbf{diag}(\max(0, \sigma - 1/t))V^T, \quad (14)$$

where $U$, $V$, $\sigma$ are from the singular value decomposition

$$\mathcal{A}(\mathbf{x}) - \mathcal{B} + Z/t = U \, \mathbf{diag}(\sigma)V^T.$$

The ADMM algorithm is summarized in Table II.

For the stopping criteria, we need to compute four values at each iteration: the primal residual $r_\text{p}$, the dual residual $r_\text{d}$, the primal tolerance $\epsilon_\text{p}$, and the dual tolerance $\epsilon_\text{d}$ [BPC+11]:

$$r_\text{p} = \mathcal{A}(\mathbf{x}) - X - \mathcal{B}$$
$$r_\text{d} = t\mathcal{A}_\text{adj}(X_\text{prev} - X)$$
$$\epsilon_\text{p} = \sqrt{pq}\epsilon_\text{abs} + \epsilon_\text{rel} \max\{\|\mathcal{A}(\mathbf{x})\|_F, \|X\|_F, \|\mathcal{B}\|_F\}$$
$$\epsilon_\text{d} = \sqrt{n}\epsilon_\text{abs} + \epsilon_\text{rel} \|\mathcal{A}_\text{adj}(Z)\|_2,$$

where $\epsilon_\text{rel}$ and $\epsilon_\text{abs}$ are the relative and absolute tolerance (for example, $\epsilon_\text{rel} = 10^{-3}$ and $\epsilon_\text{abs} = 10^{-6}$).

A few improvements can be added to the basic ADMM algorithm. Instead of using a fixed penalty parameter $t$, we can improve the convergence by updating $t$ as follows [HYW00]

$$t := \begin{cases} \tau t & \text{if} \|r_\text{p}\|_F > \mu \|r_\text{d}\|_2 \\ t/\tau & \text{if} \|r_\text{d}\|_2 > \mu \|r_\text{p}\|_F \\ t & \text{else} \end{cases}$$

with $\mu > 1$, $\tau > 1$ (for example, $\mu = 10$ and $\tau = 2$). Another improvement is to avoid the inverse in (13) by introducing an additional proximal quadratic term to the augmented Lagrangian so it cancels out the complicated quadratic term involving $\mathcal{C} + tM$ [FPST12].

Figure 1 shows a typical convergence plot of the ADMM algorithm.

## V. EVALUATION

In this section we evaluate the nuclear norm heuristic in combination with subspace identification algorithms. In our method the nuclear norm approximation is used as a pre-processing step, that computes a modified output sequence which is then passed to standard subspace system identification algorithms. For the latter purpose we have used the System Identification toolbox in MATLAB with its code n4sid. This code implements both MOESP and CVA. If the user does not specify a weighting to use, n4sid makes an automatic choice. This is the setting we have used. Also the order determination is done automatically in the code by setting order = 'best'. With this choice the model order is equal the number of singular values of $\hat{G}$ that are above the average value of the smallest and largest of the singular values in a logarithmic scale. We are always using these settings for the n4sid code whenever it is used. Specifically this is the case when we are just performing standard subspace system identification without any preprocessing. We will denote such a solution to an identification problem as the baseline solution, since this is the solution against which we will compare the nuclear norm based solutions.

A nuclear norm based solution to a system identification problem is a solution for which we have first pre-processed the output data $y_{\text{meas}}$ using a weighted nuclear norm optimization as in (11), and then applied n4sid with the settings described above as a post-processing. Different nuclear norm based solutions are obtained depending on the weighting used in (11). We denote these different solutions by referring to the corresponding weightings defined in Table I. In addition to these weightings we also consider the case when $W_1 = I$, $W_2 = I$, which we denote as NONE, and the case when we use $G$ in (4) instead of $\hat{G}$, which we denote as NOINSTR.

There are several implications from this. First of all it might be that we are using different weightings in the pre-processing stage as compared to the post-processing stage. The baseline solution might be using a third weighting. Also the order of the baseline solution and the nuclear norm solution may be different. However, neither of this matters, since we are going to compare different solutions with respect to a fit measure on the validation data.

The fit measure is the one implemented in the System Identification toolbox in the code compare, i.e.:

$$\text{fit} = 100 \left( 1 - \frac{\|y_{\text{pred}} - y\|}{\|y - \text{mean}(y)\|} \right)$$

for a single output sequence, where $y$ is the validation data output and $y_{\text{pred}}$ is the predicted output from the model. For systems with multiple outputs, we report the average of the fit. We always use different data for identification and validation. Since the solution of the pre-processing step depends on the regularization parameter $\lambda$ we have computed the fit for 20 logarithmically spaced values of the regularization parameter in the interval $2 \times 10^{-3}$ to $10^3$, and the one giving the best fit has been chosen. The parameter values $r$ and $s$ have both been set to 15 in the pre-processing step optimization, but when running the n4sid code these values have been chosen automatically by the code.

We now present evaluation results first for randomly generated models and then for the so-called DaISy collec-tion [DDDF97], an online repository of input and output data collected from real systems.

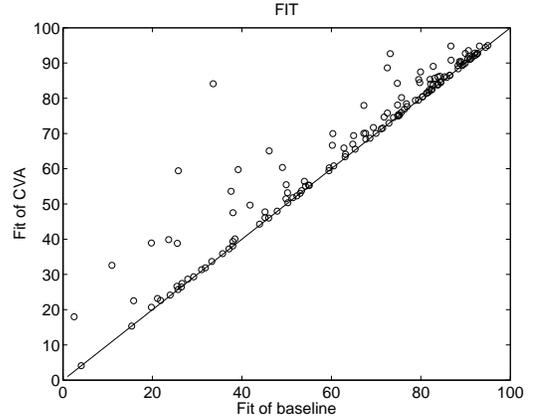

Fig. 2. Scatter plot showing the fit for the CVA-based nuclear norm method versus the fit for the baseline solution.

### A. Randomly generated models

Validation data and identification data have been generated from randomly generated state-space models. The models have been obtained with the Matlab function drss. This defines $(A, B, C, D)$. We have then set the direct term $D$ equal to zero. Moreover the Kalman gain $K$ has been generated using randn. Only single-input-single-output systems have been considered. The state dimension $n$ has been from 4 to 20 with unit steps. There has been equally many examples for each value of $n$.

The data length for identification has been 300 and the length of the validation data has been 1500. The noise $e(k)$ has been generated with randn, which means that it is white and has a standardized normal distribution with zero mean and unit covariance. The input $u(k)$ has been generated in the same way, except that it has been multiplied with a scalar $\sigma$, which has been varying from 2 to 10 with unit steps. In the experiments there have been equally many examples for each value of $\sigma$. In this way we have obtained examples with varying degree of possible fit in the data, since the signal to noise ratio has been very different in different examples.

The total number of examples considered are $(20 - 3) \times (10 - 1) = 156$. We present in Table III the percentage of cases for which each of the weightings resulted in a better fit than the baseline solution. We see that the results are best for CVA. This is also one of the fastest methods because of the dimension of the matrix. It takes about 5 seconds to compute the solution with this method. This time includes 20 runs of the optimization, 20 runs of n4sid and 20 runs of compare. Figure 2 shows a scatter plot of fits for the CVA-based nuclear norm method versus the baseline solution. It is seen that they mostly give about the same fit, but that in more than 10 % of the cases the nuclear norm approach results in a significantly better fit. Also there are no cases for which the nuclear norm solution is significantly worse than the baseline solution. The order of the solution for the nuclear norm case and for the baseline approach are different, but one cannot say that one is always greater than the other. In Figure 3 one can see that it is more common that the order of the system identified with the nuclear norm approach is higher than lower. However, very often it is the same, and when there is an increase in order it is most often only by one state. Only the cases when the fit is better with the nuclear norm approach are shown in the histogram.

TABLE III
TABLE SHOWING PERCENTAGE OF FIT BETTER THAN BASELINE SOLUTION FOR DIFFERENT WEIGHTED NUCLEAR NORM APPROACHES.

| Weight type | MOESP | N4SID | IVM | CVA | NONE | NOINSTR |
|---|---|---|---|---|---|---|
| Percentage | 82.35 | 81.70 | 90.85 | 91.50 | 76.32 | 83.66 |

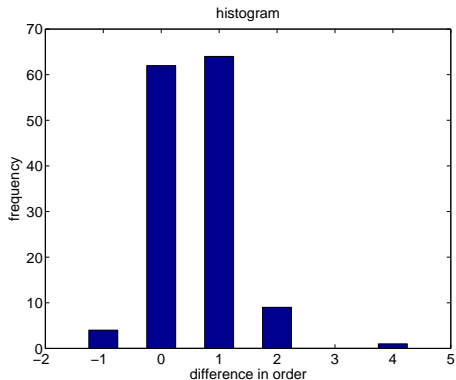

Fig. 3. Histogram for the difference in system order between the CVA-based nuclear norm solution and the baseline solution for the cases when the fit is better with the nuclear norm approach.

*B. Examples from the DaISy collection*

We have also tested the nuclear-norm based subspace system identification algorithm on ten benchmark examples from the DaISy collection. Table IV provides a brief description of the data sets. Since there is only one input-output sequence for each system, we break up the data points into two sections. The first $N_I$ data points are used in the model identification, and the next $N_V$ data points are used in the model validation. When calling the MATLAB function n4sid, we have provided additional setting nk = zeros(1,m) and focus = stability, where the former includes the estimation of the matrix $D$ and the latter forces model stability.

Table V summarizes the performance measure, fit. For clarity, the highest fit is marked with a bold font for each example in the table. The last row shows the average fit of the 10 examples for each method. We have ignored the negative fit from NONE on example 4 in the average calculation. The CVA weights gave the highest fit measure in 5 out of 10 examples, as well as in the average. The IVM and NOINSTR weighting have the next highest average fit. It is important to note that NOINSTR on example 3 and CVA on example 9 have significantly outperformed other solutions and achieved an improvement of at least 6% when comparing to the second highest fit. The MOESP weights and unweighted (NONE) formulation performed less well both in terms of average fit and in individual test cases. Generally, the results from the ten DaISy examples agree with the results obtained from the extensive simulation of random models in Section V-A.

## VI. CONCLUSIONS

In this paper a subspace identification method based on weighted nuclear norm optimization was presented. Experiments show that the use of weights in the nuclear norm approximation improves the performance in terms of fit on validation data. A second advantage is that it reduces the size of the optimization problems. The use of ADMM is also part of the speedup as compared to previous interior-point implementations. Future work is going to be devoted to tailor the ADMM algorithm to the specific application presented here.

TABLE IV

TEN BENCHMARK PROBLEMS FROM THE DAISY COLLECTION. $N_I$ IS THE NUMBER OF DATA POINTS USED IN THE IDENTIFICATION EXPERIMENT. $N_V$ IS THE NUMBER OF POINTS USED FOR VALIDATION.

|  | Data set | Description | Inputs | Outputs | $N_I$ | $N_V$ |
|---|---|---|---|---|---|---|
| 1 | 96-007 | CD player arm | 2 | 2 | 500 | 1500 |
| 2 | 98-002 | Continuous stirring tank reactor | 1 | 2 | 500 | 1500 |
| 3 | 96-006 | Hair dryer | 1 | 1 | 300 | 700 |
| 4 | 97-002 | Steam heat exchanger | 1 | 1 | 1000 | 3000 |
| 5 | 99-001 | SISO heating system | 1 | 1 | 300 | 500 |
| 6 | 96-009 | Flexible robot arm | 1 | 1 | 300 | 700 |
| 7 | 96-011 | Heat flow density | 2 | 1 | 500 | 1000 |
| 8 | 97-003 | Industrial winding process | 5 | 2 | 500 | 1500 |
| 9 | 96-002 | Glass furnace | 3 | 6 | 250 | 750 |
| 10 | 96-016 | Industrial dryer | 3 | 3 | 300 | 500 |

TABLE V

THE FIT SCORE OF SUBSPACE SYSTEM IDENTIFICATION PERFORMANCE WITH WEIGHTED NUCLEAR NORM OPTIMIZATION FOR TEN BENCHMARK PROBLEMS FROM THE DAISY COLLECTION.

|  | Baseline | MOESP | N4SID | IVM | CVA | NONE | NOINSTR |
|---|---|---|---|---|---|---|---|
| 1 | 68.4 | 72.5 | **73.5** | 71.0 | 70.0 | 72.8 | 72.3 |
| 2 | 79.5 | 80.0 | 79.6 | 82.3 | **84.8** | 75.4 | 80.9 |
| 3 | 84.2 | 84.8 | 85.0 | 84.9 | 84.9 | 84.2 | **90.2** |
| 4 | 70.8 | 70.8 | 70.8 | 70.8 | **71.8** | −245 | 70.8 |
| 5 | 82.5 | 82.5 | 82.5 | 84.4 | 84.3 | 81.9 | **84.9** |
| 6 | **96.6** | 93.1 | 92.7 | 94.8 | 95.5 | 93.5 | 96.3 |
| 7 | 83.6 | 84.8 | 83.9 | **86.2** | **86.2** | 84.8 | 85.1 |
| 8 | 58.9 | 59.5 | **62.5** | 60.2 | 60.0 | 59.8 | 59.6 |
| 9 | 55.3 | 60.9 | 56.9 | 62.0 | **66.1** | 57.9 | 57.0 |
| 10 | 40.8 | 45.0 | 45.5 | 45.2 | **47.0** | 40.8 | 42.7 |
| Average | 72.1 | 73.4 | 73.3 | 74.2 | **75.1** | 72.3 | 74.0 |